\RequirePackage{fix-cm}
\documentclass[natbib]{svjour3}                     
\smartqed  
\usepackage{graphicx}
\usepackage{epsf}
\usepackage{amssymb}
\usepackage{amsmath}
\usepackage{placeins}
\usepackage{color}

\newcommand{\apj}{Astrophys.~J.\ }
\newcommand{\apjl}{Astrophys.~J.~Lett.\ }

\newcommand{\mnras}{Mon.~Not.~R.~Astron.~Soc.\ }

\newcommand{\araa}{Annu.~Rev.~Astron.~Astrophys.\ }
\newcommand{\apjs}{Astrophys.~J.~Suppl.~Ser.\ }

\newcommand{\prl}{Phys. Rev. Lett.\ }
\newcommand{\aap}{{Astron. Astrophys.}\ }

\newcommand{\prd}{{Phys. Rev.} D}
\newcommand{\prc}{{Phys. Rev.} C}

\newcommand{\physrep}{{Phys. Rep.\ }}

\newcommand{\pasj}{Publ. Astron. Soc. Japan}
\newcommand{\apss}{Ap\&{SS}}

 \journalname{SSRv}
\newcommand{\be}{\begin{equation}}
\newcommand{\ee}{\end{equation}}
\newcommand{\beq}{\begin{eqnarray}}
\newcommand{\eeq}{\end{eqnarray}}
\newcommand\subsun[1]{{$_{\normalsize\odot}$}}

\begin{document}

\title{Crucial Physical Dependencies of the Core-Collapse Supernova Mechanism}

\titlerunning{Crucial Physical Dependencies}        

\author{A.~Burrows \and D.~Vartanyan \and J.C.~Dolence \and M.A.~Skinner \and D.~Radice}


\institute{A. Burrows \at
Department of Astrophysical Sciences, Princeton University, Princeton, NJ 08544
\email{burrows@astro.princewton.edu}
\and
D. Vartanyan \at 
Department of Astrophysical Sciences, Princeton University, Princeton, NJ 08544
\email{dvartany@astro.princeton.edu}
\and
J.C. Dolence \at
CCS-2, Los Alamos National Laboratory, P.O. Box 1663, Los Alamos, NM 87545
\email{jdoelnce@lanl.gov}
\and
M.A. Skinner \at
Livermore National Laboratory, 7000 East Ave., Livermore, CA 94550-9234
\email{skinner15@llnl.gov}
\and
D. Radice \at
Schmidt Fellow, Institute for Advanced Study, 1 Einstein Drive, Princeton, NJ 08540
\email{dradice@astro.princeton.edu}
}

\date{Received: October 24, 2017 / Accepted: November 20, 2017}

\maketitle

\begin{abstract}
We explore with self-consistent 2D F{\sc{ornax}} simulations
the dependence of the outcome of collapse on many-body
corrections to neutrino-nucleon cross sections, the nucleon-nucleon bremsstrahlung rate, 
electron capture on heavy nuclei, pre-collapse seed
perturbations, and inelastic neutrino-electron and neutrino-nucleon
scattering. Importantly, proximity to criticality amplifies
the role of even small changes in the neutrino-matter couplings,
and such changes can together add to produce outsized effects.
When close to the critical condition the cumulative result of
a few small effects (including seeds) that individually have only modest consequence
can convert an anemic into a robust explosion, or even a dud into a blast.
Such sensitivity is not seen in one dimension and may explain
the apparent heterogeneity in the outcomes of detailed simulations
performed internationally. A natural conclusion is that the
different groups collectively are closer to a realistic understanding of
the mechanism of core-collapse supernovae than might have seemed apparent.
\end{abstract}


\section{Introduction}
\label{intro}

A goal of core-collapse supernova theory is to explain the mechanism of explosion.
It is an accepted truism of the field that a necessary condition for this is
that complicated multi-dimensional simulation codes incorporating the requisite
neutrino, nuclear, and gravitational physics reproduce such explosions robustly,
yielding at the very least the requisite asymptotic energies, residual neutron star masses, 
and nucleosynthesis.  A simple analytic explanation has not been forthcoming and, given
the manifest complexities of the process, would not be deemed credible.
It is thought that one litmus test of success would be such a demonstration 
for a non-rotating model between $\sim$8 and $\sim$20 M$_{\odot}$, the progenitor 
ZAMS (Zero-Age Main-Sequence) mass regime that must provide the lion's share of core-collapse supernova 
events.

However, to date the various groups engaged in such efforts around the world 
have failed to agree on the outcomes of the collapse of otherwise similar progenitor massive 
star cores. This is despite claims to have embedded the necessary physics and microphysics
into the simulations. In fact, the numerical algorithms, resolutions, and input physics 
all differ, and even when the physics is deemed similar, the implementations and 
approximations surely differ.  The ORNL group (Bruenn et~al. 2013,2016), with 
their CHIMERA code, uses multi-group flux-limited diffusion 
(MGFLD) neutrino transport (Bruenn 1985), the VH-1 Newtonian hydrodynamics package, a monopole 
correction for general relativity (GR) (Marek et al. 2006), but multiple one-dimensional 
solves for multi-dimensional transport using the so-called ``ray-by-ray+" approach (Buras et al. 
2003; Burrows, Hayes, \& Fryxell 1995). Such a dimensional reduction for the transport, 
particularly manifest in 2D, ignores lateral, non-radial radiative transport, which has 
been shown to be of quantitative (Ott et al. 2008; Brandt et al. 2011; 
Burrows 2013; Dolence, Burrows, \& Zhang 2015; Sumiyoshi et al. 2015)
and qualitative (Skinner, Burrows, \& Dolence 2016) importance.  
To the point, Skinner et al. (2016) have shown that the ray-by-ray anomalies in the 
angular distribution of the radiation field and corresponding neutrino 
heating rates can reinforce the axial sloshing motions in 2D 
and push the shock into explosion\footnote{However, the general
absence in 3D of either an axial effect or the pronounced sloshing seen in
2D may be rendering 3D simulations performed with the ray-by-ray approach
less problematic. This has yet to be tested.}. Bruenn et al. (2013,2016) find explosions in 2D (axially 
symmetry) for all the progenitors they studied (the 12-, 15-, 20-, and 25-M$_{\odot}$ 
models of Woosley \& Heger 2007), but the models all explode at about the same 
post-bounce time ($\sim$100 milliseconds) and the shock radii never decrease in value.
When performed in 3D for the 15-M$_{\odot}$ model of Woosley \& Heger (2007), 
Lentz et al. (2015) obtain a weaker explosion delayed 
in its onset by an extra $\sim$100 milliseconds.

M\"uller et~al. (2012ab), using the conformally-flat CoCoNuT hydrodynamics 
code in combination with the VERTEX transport solver and the ray-by-ray+ approach, 
find that models explode in 2D, but explode with lower energies and 
at post-bounce explosion times that differ significantly from those found 
by the ORNL group. In addition, their shock radii generally decrease from 
a peak occurring near a time $\sim$200 ms after bounce, before 
increasing just prior to explosion hundreds of milliseconds later
(and all at different post-bounce times). Using the VERTEX-PROMETHEUS 
code, the Garching group obtained weak explosions in 2D for a 11.2-M$_{\odot}$ 
progenitor (Buras et al. 2006) and for a rotating 15-M$_{\odot}$ 
progenitor (Marek \& Janka 2009), though the outcomes when proper correction is made
for the corresponding mantle binding energies were not clear.  More recently, the Garching 
group (Summa et al. 2016) obtained explosions in 2D for a broad range of Woosley \& Heger 
(2007) progenitors from 11 to 28 M$_{\odot}$, but these calculations as well did not comport 
with those of the ORNL group in the timescales and energies seen. The Garching group 
has also used for all their multi-D VERTEX supernova simulations a variant of the 
problematic ray-by-ray+ dimensionally-reduced transport approach.  

However, Murphy, Dolence, \& Burrows (2013) and                  
Couch \& Ott (2015), and originally Burrows, Hayes, \& Fryxell (1995),
highlight the importance of turbulent pressure behind the shock as an
aid to explosion, but note that such a pressure is larger (artificially)
in 2D than in 3D.  Turbulent pressure is also (possibly) more anisotropic in 2D, favoring the
radial stress component, further enhancing the prospects (again, artificially) for overcoming the
accretion ram pressure. Hence, the current explosions in 2D may in part, or at times, be numerical
artifacts.

Importantly, though Hanke et al. (2012) have speculated that the sloshing motion 
oftimes associated in 2D with the SASI (standing accretion shock instability) may 
be crucial to explosion, the recent non-rotating default VERTEX-PROMETHEUS calculations in 3D by that same 
Garching group (that at times manifest the SASI) do not explode (Hanke et al. 2013; Tamborra et al. 2014), even 
though the corresponding 2D simulations did. As an aside, Burrows et al. (2012) note that
such a pronounced axial sloshing motion is rarely in evidence in 3D.  
However, with altered microphysics, in particular with a change in the axial-vector
coupling constant ($g_A$) due to a speculative enhanced effect on the nucleon spin 
of the strange quark, Melson et al. (2015) do obtain a weak explosion in 3D of the 
Woosley \& Heger (2007) 20-M$_{\odot}$ in 3D. The difference in outcome is due to 
the consequent decrease in the neutrino-neutron neutral-current scattering cross section in 
the neutron-rich envelope of the proto-neutron star bounded by the stalled shock
and the resultant increase in the electron-neutrino luminosity and average energy 
that are instrumental in heating this envelope.  However, the magnitude of the 
strangeness correction employed by Melson et al. may be larger than experiment
allows (Ahmed et al. 2012; Green et al. 2017). Furthermore, it is not clear that all these currently published 
under-powered 3D explosions will actually succeed as an explosion after tranversing 
the entire star and after the mass cut between ejecta and residual core
is determined hydrodynamically. An exception is the recent work of M\"uller et al. (2017),
who perform 3D multi-group simulations using their simpler FMT method (M\"uller \& Janka 2015)
for an 18-M$_{\odot}$ progenitor model evolved in 3D to collapse.  Such a 3D progenitor
naturally manifests seed perturbations that M\"uller et al. (2017) show
are instrumental in leading to an explosion with a respectable energy.

The calculations of Burrows et al. (2006,2007a) and Ott et al. (2008),
using the VULCAN/2D code, and Dolence, Burrows, \& Zhang (2015), using the
CASTRO code (Zhang et al. 2011,2013) employed multi-dimensional transport,
and not ray-by-ray+, but in neither of these 2D studies did the authors see
explosions by the neutrino mechanism. VULCAN/2D did not have all the terms to
order $v/c$ in the transport, but CASTRO did, and the results were similar (Dolence,
Burrows \& Zhang 2015). Importantly, neither VULCAN/2D nor CASTRO made corrections for the
effects of general relativity, and this could explain in part the different outcome.
However, since other self-consistent calculations (summarized previously)
that obtained explosions in 2D used the ray-by-ray scheme, while the VULCAN/2D
and CASTRO studies did not, one is tempted to suggest that the ray-by-ray
approach may in 2D be yielding qualitatively incorrect results.

Suwa et al. (2010) obtain an explosion in 2D of a 13-M$_{\odot}$ progenitor, while Takiwaki et al. (2012) obtain
an explosion for an 11.2-M$_{\odot}$  progenitor in both 2D and 3D. Both these efforts, however, neglect 
$\nu_{\mu}$ and $\nu_{\tau}$ neutrinos, which constitute roughly 50\% of the total neutrino losses after bounce,
and use the IDSA (Liebend\"orfer et al. 2009) and ray-by-ray approximations for the transport. 
{Suwa et al. (2016) continued their explorations in 2D with a large suite
of progenitor simulations, but continue to neglect $\nu_{\mu}$ and $\nu_{\tau}$ neutrinos and 
use the IDSA plus ray-by-ray+ transport methods. They highlight a possible role in the 
explosion systematics of transitions in the mass accretion rate.} 
Ott et al. (2013) perform fully general-relativistic 3D simulations, employing
a leakage scheme for all the neutrinos.  Their goal was to explore the relative role 
of the SASI and neutrino-driven convection, and they found for their simulation of 
the 27-M$_{\odot}$ progenitor of Woosley, Heger, \& Weaver (2002) that 
neutrino-driven convection dominated  once it started.
Roberts et al. (2016), using full GR hydrodynamics and
an M1 transport scheme (\S\ref{method}) roughly similar to ours (but without the velocity-dependent
terms in the transport, inelastic scattering, or many-body effects), emphasize
the importance of spatial resolution in determining whether and how the same 27-M$_{\odot}$ model explodes, as well as
the different outcomes for octant and full $4\pi$ steradian simulations. Kuroda et al. (2016)
have achieved a fully general-relativistic 3D code, also using the M1 transport closure, but have yet to
simulate progenitor models beyond a few tens of milliseconds post-bounce.  Pan et al. (2016)
have constructed a 2D non-relativistic (Newtonian) neutrino radiation/hydrodynamic scheme using the IDSA
transport approach for the $\nu_e$ and $\bar{\nu}_e$ neutrinos and a leakage scheme
for the $\nu_{\mu}$ neutrinos. Importantly, they do not use the ray-by-ray+ approach,
but follow the transport of $\nu_e$ and $\bar{\nu}_e$ neutrinos multi-dimensionally.
They obtain explosions for all the progenitor models studied.

Suwa et al. (2010) and Nakamura et al. (2014) found that rotation aided explosion, mostly by
rotationally expanding the size of the gain region and increasing the mass it contained.
Iwakami, Nagakura, \& Yamada (2014) and Takiwaki et al. (2016) highlight the rotational excitation of $m=1$
spiral-arm modes and find a role for non-axisymmetric rotational instabilities.  {Iwakami, Nagakura, \& Yamada (2014)
used a light-bulb neutrino scheme, and neglected ``$\nu_{\mu}$" neutrinos.} 
Both Nakamura et al. (2014) and Takiwaki et al. (2016) observed equatorial explosions in the rapidly-rotating context.
Recently, Janka et al. (2016)  and Summa et al. (2018) published a rapidly-rotating 
3D model that exploded and highlighted the potential role of an ``$m=1$" structure in the gain region.
Earlier, Fryer \& Heger (2000) used SPH for the hydrodynamics and a simplified gray
scheme for the neutrino transport to explore the role of rapid rotation.  
Moreover, in all of these studies the initial rotation rate was not only high in the mantle, but was high in the core.
Such rapid initial spins (periods of a few seconds) and final spins (periods of 
$\sim$2$-$10 milliseconds) seem not to be consistent with inferred pulsar
spin periods at birth (crudely, $\sim$300 $\pm$ 200 milliseconds; Emmering \& Chevalier 1989;
Faucher-Giguere \& Kaspi 2006; Popov \& Turolla 2012; Noutsos et al. 2013)
and may be associated only with hypernovae (Burrows et al. 2007c) and/or gamma-ray bursts
(MacFadyen \& Woosley 1999). 

In this paper, we explore, with self-consistent 2D F{\sc{ornax}} (\S\ref{method}) simulations, the dependence 
of the outcome of collapse (most notably whether the model explodes) on neutrino-nucleon scattering rates
(via modifications of in-medium response corrections due to many-body effects), pre-collapse convective 
perturbations, inelastic neutrino-electron scattering, and inelastic neutrino-nucleon scattering. 
We also continue our study, started in Skinner et al. (2016), of the issues raised by 
the use of the ray-by-ray+ method. What we find is that when the proto-neutron star 
bounded by a stalled shock is close to the critical condition for explosion 
(Burrows \& Goshy 1993), as it easily can be in the turbulent multi-D context, the sensitivity 
to explosion of small changes in the physical inputs is amplified.  The magnitude 
of such changes might be only $\sim$20\%, but the result can be qualitatively different, 
in particular whether the model explodes.  In the 1D (spherical) case, the core is not sensitive to 
comparable changes (``Mazurek's Law"), but in the multi-D turbulent (and chaotic) 
context, small changes in the physics can have a qualitative effect on explodability.  
This may explain why the various groups around the world simulating core-collapse 
supernovae can witness very different outcomes, despite the fact that they ostensibly 
are incorporating almost the same microphysics and similar computational approaches. 
Small differences are amplified near criticality in this chaotic context. 
The availability of a new generation of fast, but accurate, simulation codes, 
such as F{\sc{ornax}} (\S\ref{method}), and significant supercomputer 
resources enable rapid multi-parameter investigations in the multi-dimensional (in particular 2D), 
multi-physics context. Such wide-ranging explorations reveal patterns not easily 
discerned when one (or only a few) simulations are the focus of a paper and its (or their) results 
solely are mined.

\section{Numerical Method and Computational Setup}
\label{method}

We have developed an entirely new multi-dimensional, multi-group radiation/hydrodynamic code,
F{\sc{ornax}}, for the study of core-collapse supernovae.  This code
is described in detail in an upcoming paper (in preparation).  
For the purposes of this paper, we note that it employs spherical coordinates in 
one and two spatial dimensions, solves the comoving-frame, multi-group, 
two-moment, velocity-dependent transport equations to O($v/c$), 
and uses the M1 tensor closure for the second and third moments of the radiation
fields (Dubroca \& Feugeas 1999; Vaytet et al. 2011). Three species of neutrino 
($\nu_e$, $\bar{\nu}_e$, and ``$\nu_{\mu}$" [$\nu_{\mu}$, $\bar{\nu}_{\mu}$, 
$\nu_{\tau}$, and $\bar{\nu}_{\tau}$ lumped together]) are followed using 
an explicit Godunov characteristic method applied to the radiation transport operators, but an implicit solver 
for the radiation source terms.  

The hydrodynamics in F{\sc{ornax}} is based on a directionally-unsplit Godunov-type finite-volume 
method.  Fluxes at cell faces are computed with the fast and accurate HLLC approximate 
Riemann solver based on left and right states reconstructed from the underlying volume-averaged 
states.  In the interior, to alleviate Courant limits due to converging angular zones, 
the code can deresolve in angle with decreasing radius, conserving hydrodynamic and radiative
fluxes in a manner similar to the method employed in AMR codes at refinement boundaries.
Gravity is handled in 2D and 3D with a monopole or a multipole solver (M\"uller \& Steimetz 1995).
When using the latter, we generally set the maximum spherical harmonic order necessary 
equal to twelve. The monopole gravitational term is altered to accommodate 
approximate general-relativistic gravity (Marek et al. 2006), 
and we employ the metric terms, $g_{rr}$ and $g_{tt}$, derived from this potential 
to incorporate general relativistic redshift effects in the neutrino 
transport equations (in the manner of Rampp \& Janka 2002).  
In 2D, rotation and a third component of the velocity vector can be included 
in the hydrodynamics. We use the SFHo equation of state (EOS) by default (Steiner et al. 2013),
but in this paper do compare with results using the DD2 EOS (Banik et al. 2014) \footnote{In a 
earlier version of this paper, we used the K = 220 MeV Lattimer-Swesty 
equation of state (Lattimer \& Swesty 1991) for all our simulations, but 
since this EOS has been shown to be inconsistent with recent measurements we 
redid the models and paper using EOSes still consistent with what is known.}.

For these simulations, we follow twenty energy ($\varepsilon_{\nu}$) groups 
for each of the $\nu_e$, $\bar{\nu}_{e}$, and ``$\nu_{\mu}$" species {(again, 
the four species, $\nu_{\mu}$, $\bar{\nu}_{\mu}$, $\nu_{\tau}$, and 
$\bar{\nu}_{\tau}$ lumped together.)}  For the $\nu_e$ types, the neutrino 
energy $\varepsilon_{\nu}$ varies logarithmically from 1 MeV to 300 MeV, while it varies
from 1 MeV to 100 MeV for the $\bar{\nu}_{e}$s and $\nu_{\mu}$s. We have performed calculations 
with forty energy groups and found little difference in the results.  
The radial coordinate, $r$, runs from 0 to 20,000 kilometers (km) in 608 zones.
The radial grid smoothly transitions from uniform spacing with $\Delta r=0.5$ km
in the interior to logarithmic spacing, with a transition
radius near $\sim$100 km. The polar angular grid spacing covers the full $180^\circ$ and varies smoothly from
$\approx 0.95^\circ$ at the poles to $\approx 0.65^\circ$ at the equator in 256 zones. 

A comprehensive set of neutrino-matter interactions are followed in F{\sc{ornax}},
and these are described in Burrows, Reddy, \& Thompson (2006). 
Inelastic neutrino-nucleon scattering is handled using a modified version
of the Thompson, Burrows, \& Pinto (2003) approach (\S\ref{inelastic}). 
Our neutrino microphysics is more comprehensively listed in our upcoming code paper.

\section{Many-Body Neutrino Response Corrections}
\label{horo}

Melson et al. (2015) invoked a modification in the axial-vector coupling constant ($g_A$)
due to a possible strangeness contribution to the nucleon spin of $g_A^s = -0.2$.  This results in an
approximate decrease in the neutrino-nucleon scattering rate of $\sim$20\% and in Melson et al. 
the consequence was an explosion in 3D, even though they did not witness an explosion 
in 3D when using their default microphysical suite.  Curiously, without their strangeness correction, 
the same model exploded in 2D. However, the value of the correction, $g_A^s$, they employed in their 3D model
is likely too large and $g_A^s$ may be closer to zero (Ahmed et al. 2012; Green et al. 2017).

Many-body corrections to neutral-current and charged-current neutrino-nucleon 
interactions have been discussed in the past (Burrows \& Sawyer  1998,1999; 
Hannestad \& Raffelt 1998; Reddy et al. 1999; Roberts et al. 2012) in the context 
of proto-neutron stars and supernovae, and have been
known to affect the neutrino-matter reaction rates.  Burrows \& Sawyer (1998) in 
particular suggested that many-body corrections to the axial-vector
and vector structure factors for neutrino-nucleon scattering could be of a magnitude sufficent
to be of relevance to the viability of the neutrino-driven mechanism of core-collapse supernovae, but
did not provide a robust estimate of the magnitude of this diminution much
below nuclear densities. 

The effect of the many-body correction to neutral-current neutrino-nucleon scattering
in the supernova context is mostly due to the increase in the
$\nu_{\mu}$ luminosity occasioned by the decrease in the associated
$\nu_{\mu} + (n,p) \rightarrow \nu_{\mu} + (n,p)$ scattering cross section for densities
above $\sim$10$^{12}$ g cm$^{-3}$; this causes a further compression in the core.  Such
a compression, similar to the effect of GR, raises the temperatures near the $\nu_e$ and $\bar{\nu}_{e}$
neutrinospheres, thereby raising their associated luminosities and average emergent
neutrino energies. These changes increase the neutrino-matter heating rates in
the gain region and, hence, facilitate explosion. Since the super-allowed charged-current
absorption reactions still dominate the $\nu_e$/$\bar{\nu}_{e}$-matter interaction rates, the direct effect of this
structure factor correction to the axial-vector term in the neutrino-nucleon scattering rate
on the $\nu_e$ and $\bar{\nu}_{e}$ luminosities is slight.

A many-body structure factor ($S_A$) correction to the axial-vector term in the
neutrino-nucleon scattering rate due to the neutrino response to nuclear matter at (low)
densities below $\sim$10$^{13}$ g cm$^{-3}$ was recently derived by Horowitz et al. (2017)
using a virial approach.
{These authors derive a fit:
\begin{equation}
S_A = \frac{1}{1 + A(1+Be^{-C})}\, ,
\end{equation}
where
\begin{align}
A = A_0\frac{n(1-Y_e + Y_e^2)}{T^{1.22}}\\
B= \frac{B_0}{T^{0.75}}\\
C=C_0\frac{nY_e(1-Y_e)}{T^{0.5}} + D_0\frac{n^4}{T^6}\, .
\end{align}
In these equations, $T$ is the temperature in MeV, $Y_e$ is the electron fraction,
$n$ is the baryon density in fm$^{-3}$, $A_0 = 920$,
$B_0 = 3.05$, $C_0 = 6140$, and $D_0 = 1.5 \times 10^{13}$.}
Horowitz et al. (2017) join their fit to
Burrows \& Sawyer (1998) for the higher densities, but
were most careful fitting their formula for temperatures
between 5 and 10 MeV. {Nevertheless, though Horowitz et al. (2017)
intended their fit to apply at all densities, temperatures, and Y$_e$s,
(and we use it in this paper at all thermodynamic points), one should be aware
that no current approach to this physics is likely to be correct
at the highest densities above $\sim$$10^{14}$ g cm$^{-3}$.
Therefore, use of this formula in supernova and proto-neutron-star cores
should be considered provisional. Fortunately, since such densities are too
high to affect the evolution during the first second post bounce,
and are most relevant during the later proto-neutron-star cooling phase, the
values of the Horowitz correction at the highest densities are not germane to
the conclusions of this paper.}

However, many-body effects in
the charged-current sector and on absorption may be comparable
(Burrows \& Sawyer 1999; Roberts et al. 2012; Fischer 2016), but
have not yet been factored in.  Therefore, neglected is the effect of final-state
nucleon blocking for charged-current absorption reactions.
Blocking is important only at high densities, where many-body
interaction effects are likely to be even larger.  Therefore,
we have postponed their inclusion until such corrections, which
must for self-consistency be done with the same interaction model that
informed the equation of state employed, are available.

The fit derived by Horowitz et al. (2017) to the structure factor applied 
to $g_A^2$ translates into a decrease in the neutrino-nucleon scattering 
cross section in the crucial region at and deeper than the various 
neutrinospheres of $\sim$5\% to $\sim$35\%. This effect is a function of 
density, temperature, and electron fraction ($Y_e$). Horowitz et al. (2017) state  
that the corresponding structure factor for the vector current is likely greater 
than one, but since the vector contribution to neutrino-nucleon scattering is 
small, this is likely to be subdominant.  The upshot is a potentially important,
and physically plausible, decrease in the neutrino-matter scattering rates that translates into 
an increase in the driving $\nu_e$ and $\bar{\nu}_{e}$ luminosities and average energies, thereby
increasing the heating rate in the gain region. 

We note that the corresponding many-body correction for charged-current interactions may be in 
the same direction (Fischer 2016; Burrows \& Sawyer 1999, although see Roberts et al. 2012) and 
could also be important (\S\ref{brems}).  The correction would be small at low densities in the gain region, but 
higher deeper inside, where the neutrinospheres reside. If the rates are suppressed, this 
could increase the $\nu_e$ and $\bar{\nu}_{e}$ luminosities, while simultaneously 
not decreasing the heating in the gain region and is not like a uniform correction 
at all radii.  Such behavior, if it obtains, would be near optimal for aiding 
the explosion.

\section{Inelastic Scattering}
\label{inelastic}

Neutrino-electron scattering rates and cross sections are much smaller than
those for neutrino-nucleon scattering, which themselves are smaller still than
those for super-allowed charged-current reactions such as $\nu_e + n \rightarrow p + e^-$.
For $\varepsilon_{\nu} = 10$ MeV, this deficit is approximately two orders of magnitude.  However,
due to the small mass of the electron, the energy transfer to the matter during a
``Compton-like" neutrino-electron scattering is on average quite large, while, due
to the large mass of the nucleon, the corresponding average energy transfer during
neutrino-nucleon scattering is rather small.  Therefore, the large cross section for
neutrino-nucleon scattering multiplied by the small associated energy transfer can be comparable to the
product of the small neutrino-electron cross section with the large per-interaction
energy transfer and depend upon temperature, density, and neutrino energy
(Janka et al. 1996; Thompson, Burrows, \& Horvath 2000). The upshot is that
inelastic scattering off both electrons and nucleons can modify thermal
profiles and heating rates exterior to the neutrinospheres and in the
gain region and contribute to explosion by the neutrino heating mechanism.
Moreover, M\"uller et al. (2012b) make the point that heating by $\nu_{\mu}$-nucleon
inelastic energy transfer can boost the temperatures near the
$\nu_e$ and $\bar{\nu}_{e}$ neutrinospheres and results in slightly higher $\nu_e$ and $\bar{\nu}_{e}$
energy luminosities, which in turn enhance heating behind the shock correspondingly.

The detailed theory of the structure functions and redistribution kernels
for such inelastic scattering, including the effects of final-state blocking
and the thermal spectrum of the targets, can be found in Burrows, Reddy, \&
Thompson (2006), Reddy et al. (1999), Thompson, Burrows, \& Pinto (2003), and Thompson, Burrows,
\& Horvath (2000). Pioneering work on inelastic scattering off electrons in the core-collapse context
can be found in Bruenn (1985) and Mezzacappa \& Bruenn (1993). However, those latter papers
were focussed on the downscattering effect due to inelastic $\nu_e$-$e^-$ scattering on
the electron neutrinos produced during infall and the consequent decrease in the trapped
lepton fraction. Since a larger trapped lepton fraction could help facilitate
immediate post-bounce explosions (Burrows \& Lattimer 1983), this quantity was more
relevant when the prompt hydrodynamic supernova mechanism still seemed viable.
However, with the emergence of the delayed, neutrino-driven mechanism (Bethe \& Wilson 1985),
and the conclusion that the prompt mechanism could not work due to catastrophic
neutrino losses at and around shock breakout, the value of the trapped lepton
fraction, and its precise value, receded in significance.

Nevertheless, heating behind the stalled shock due to inelastic energy transfer
from neutrinos to both electrons and nucleons, or boosting the $\nu_e$
and $\bar{\nu}_{e}$ luminosities by $\nu_{\mu}$ downscattering near their
neutrinospheres (M\"uller et al. 2012b), may help achieve the critical condition
for explosion, particularly when in concert with the inclusion of the in-medium neutrino response
(Horowitz et al. 2017; \S\ref{horo}) and the slightly net positive influence of GR.
Heating by inelastic neutrino-electron scattering is still sub-dominant with respect to that due to
charged-current $\nu_e$ and $\bar{\nu}_{e}$ absorption, and in 1D (spherical) simulations
there is almost no hydrodynamic consequence of its inclusion (Thompson, Burrows, \& Pinto 2003).
The same can be said of inelastic neutrino-nucleon scattering (Janka et al. 1996;
Burrows \& Sawyer 1998,1999; Thompson, Burrows, \& Horvath 2000).  However, in
the realistic multi-D context of the core-collapse phenomenon, the cumulative
effect of the addition of a few sub-dominant heating mechanisms, compression
due to enhanced $\nu_{\mu}$ neutrino losses and consequent $\nu_e$ and $\bar{\nu}_{e}$
neutrinosphere heating\footnote{as in neutrino-electron scattering and many-body
scattering rate suppression}, and greater proximity to the critical condition due to multi-D
effects\footnote{Examples include the enhancement of the stress behind the shock due to turbulent pressure
(Burrows, Hayes, \& Fryxell 1995; Murphy, Dolence, \& Burrows 2013; Janka 2012; Burrows 2013)
and the modest increase in the dwell time in the gain region of the post-shock matter
(Murphy \& Burrows 2008; Dolence et al. 2013)} amplifies the leverage of even small additions to
the gain-region heating power over their effects individually and can convert an anemic
explosion into a robust explosion, or even a dud into a blast. One recalls that
the original delayed mechanism of Wilson required for explosion only a modest enhancement of $\sim$25\%
in the neutrino luminosity\footnote{In that case, it was due to ``neutron-finger convection,"
subsequently later shown not to occur (Bruenn \& Dineva 1996; Dessart et al. 2006).}.

From the work of Thompson, Burrows, \& Horvath (2000) and Tubbs (1979),
we find that the crossover energy between upscattering and downscattering
is nearer 6$k_B T$ (not 3$k_B T$, as in M\"uller \& Janka 2015), where
$k_B$ is Boltzmann's constant and $T$ is the temperature, around densities
of $\sim$10$^{11}$ g cm$^{-3}$ to $\sim$10$^{13}$ g cm$^{-3}$ and so
our approximate redistribution rate for $\nu_{\mu}$s is proportional to
$\kappa_{scat}(\varepsilon_{\mu} - 6 k_BT)/m_nc^2$, where $\kappa_{scat}$
is the $\nu_{\mu}$ scattering opacity and $m_n$ is the neutron mass.

Inelastic scattering off nucleons
lowers the $\nu_{\mu}$ luminosity, while the corresponding
quantities for the $\nu_e$ and $\bar{\nu}_{e}$ are slightly increased (M\"uller et al. 2012b).  The
latter responses explain the positive effect on explodability
of the inclusion of inelastic scattering off nucleons. The direct effect
of such inelasticity is in the core, and the
effect in the shocked mantle is indirect.

\section{Microphysical Dependences of the 2D Explosion Models}
\label{role}

For a set of 4 fiducial models (with progenitor masses of 13, 15, 16, and 20 M$_{\odot}$), 
we performed a parameter study of input physics highlighting the role of 
small changes in promoting (or preventing) explosion. We
refer the reader to the upcoming paper by Vartanyan et al. (2018, in preparation) 
for a more detailed description of these simulations. When models 
explode, the time to explosion can vary significantly with inputs.  The relative 
time of explosion can help one gauge the role of the inputs in question
\footnote{The ``time of explosion" is sometimes defined as the time the mean
shock radius achieves 400 km.  We think this definition somewhat arbitrary, but acknowledge
some degree of arbitrariness in any definition, however useful.  We here
define the time of explosion as the approximate time at which the curve of the mean shock radius
versus time is inflected upwards.  All our models that inflect in this way explode
and all these models have been continued to at least one second after bounce. The
mean shock radii of these exploding models all achieve radii beyond 5000 kilometers by 
the end of the simulation.}.
While there are many combinatoric possibilities, we settled on just a few      
comparisons to demonstrate the effects we see universally.
We employ the notation IES, INS, MB, pert, and rbrp to indicate ``inelastic neutrino-electron,"
``inelastic neutrino-nucleon," ``many-body," ``perturbations,", and ``ray-by-ray+," respectively.

\subsection{Inelastic Scattering}
\label{inel}

\begin{figure*}
\includegraphics[width=\textwidth, angle = 0]{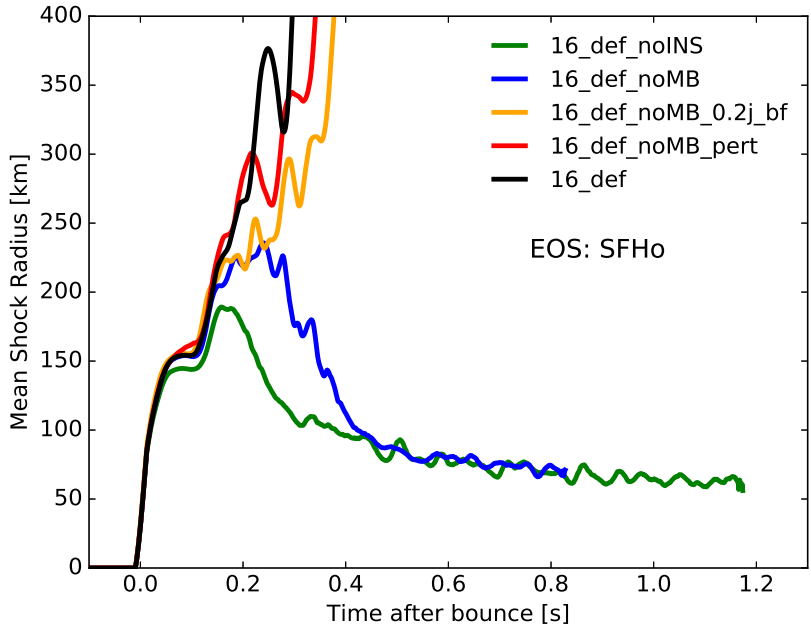}
\caption{Shock radii (in km) versus time post-bounce (in s) for variations 
on the default 16 M$_{\odot}$ progenitor. In all figures, default model ``def" 
refers to inclusion of inelastic scattering off both electrons (IES) and nucleons (INS), 
as well as the inclusion of the Horowitz et al. (2017)
many-body correction (MB). This model explodes at $\sim$250 ms post-bounce. We 
then remove and add certain inputs, denoted by a subscript with ``def". 
Removing either the many-body correction (blue, ``16$_{\_}$def$_{\_}$noMB") or 
inelastic scattering off nucleons (green, ``16$_{\_}$def$_{\_}$noINS") leads to a dud. 
However, even without the many-body correction (noMB), adding either 
perturbations (red) ( ``$_{\_}$pert"; \S\ref{perturb}) or modifying the opacity table 
to include Fischer's (2016) correction to the nucleon-nucleon bremsstrahlung rate 
(``bf") and only 20\% of the electron capture rate (Juodagalvis et al. 2010) 
on heavy nuclei (orange, ``0.2j"), leads to an explosion $\sim$50 and $\sim$100 ms, 
respectively, after our default model. This helps illustrate the sensitive dependence 
of the outcome $-$ explosion or dud $-$ on the microphysical inputs when near criticality.}
\label{fig:rshock_full}
\end{figure*}

Models with only inelastic scattering off electrons (IES),
inelastic scattering off both electrons and nucleons (IES$_{\_}$INS), and
additionally with the many-body effect (IES$_{\_}$INS$_{\_}$MB) can illuminate
the respective roles of each \footnote{In a previous version of this paper, 
we had employed an opacity file that was compromised by a compiler bug.
The net result of this error was a slightly enhanced magnitude
of the neutral-current many-body effect. This error has now been fixed.}.  
Figure \ref{fig:rshock_full} compares outcomes for the 16-M$_{\odot}$ 
progenitor. We toggled the role of inelastic scattering off electrons and 
nucleons as well as the many-body effect for the 16 M$_{\odot}$ progenitor of Woosley \& Heger (2007), which 
we found to explode early at $\sim$300 ms using the default setup IES$_{\_}$INS$_{\_}$MB. 
However, performing the simulation with any of these components removed prevents explosion.

\begin{figure*}
\includegraphics[width=\textwidth, angle = 0]{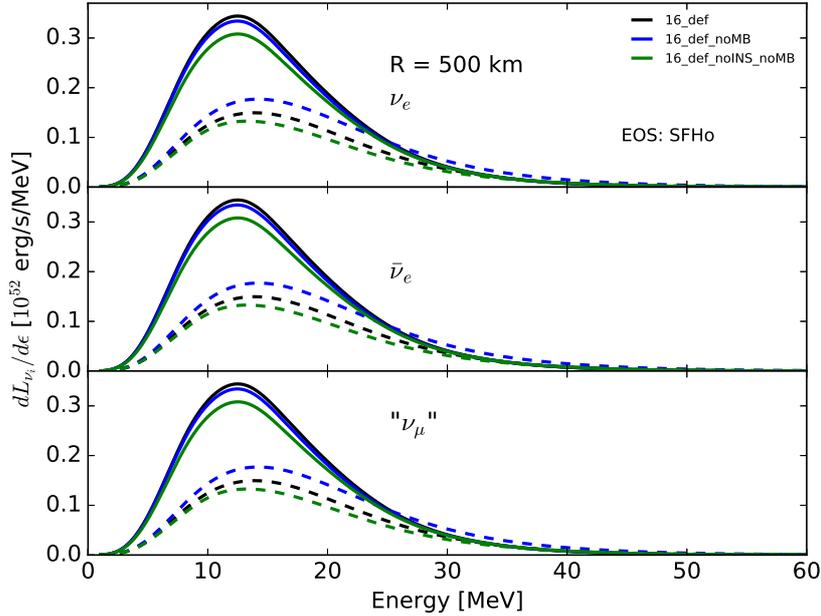}
\caption{The role of inelastic scattering off electrons and nucleons and the
neutral-current many-body correction on the emergent spectra (in 10$^{52}$ erg/s/MeV)
at 500 km at 100 (solid) and 400 (dashed) ms post-bounce. At early times, 
prior to explosion, both inelastic scattering off electrons and 
nucleons and the many-body correction lead to upscattering. At 400 ms, 
the default model has exploded and hence has a diminished spectrum
vis-\`a-vis the non-exploding model 16\_def\_noMB.}
\label{fig:spectra_micro}
\end{figure*}

\begin{figure*}
\includegraphics[width=0.5\textwidth, angle = 0]{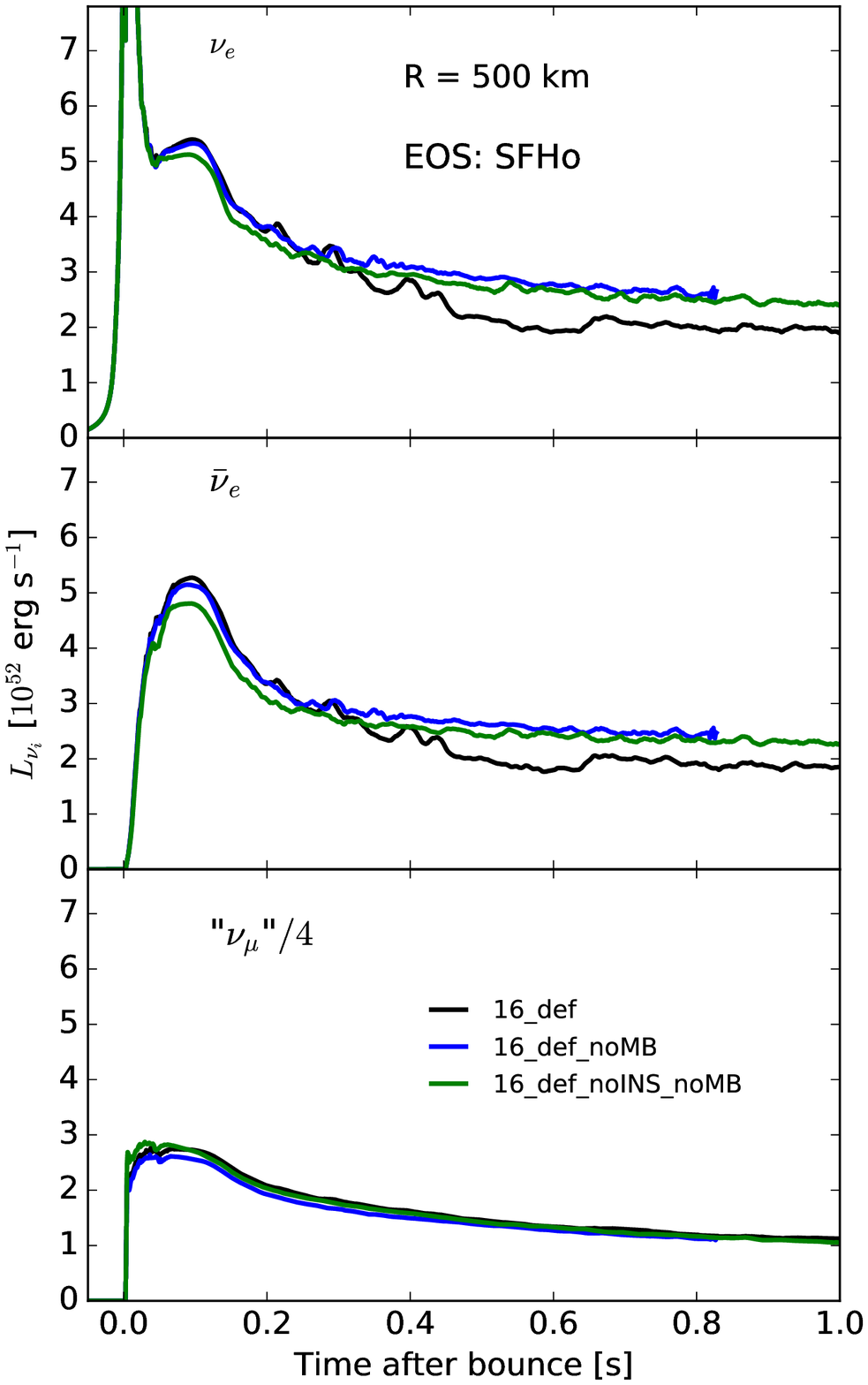}
\hfill
\includegraphics[width=0.5\textwidth, angle = 0]{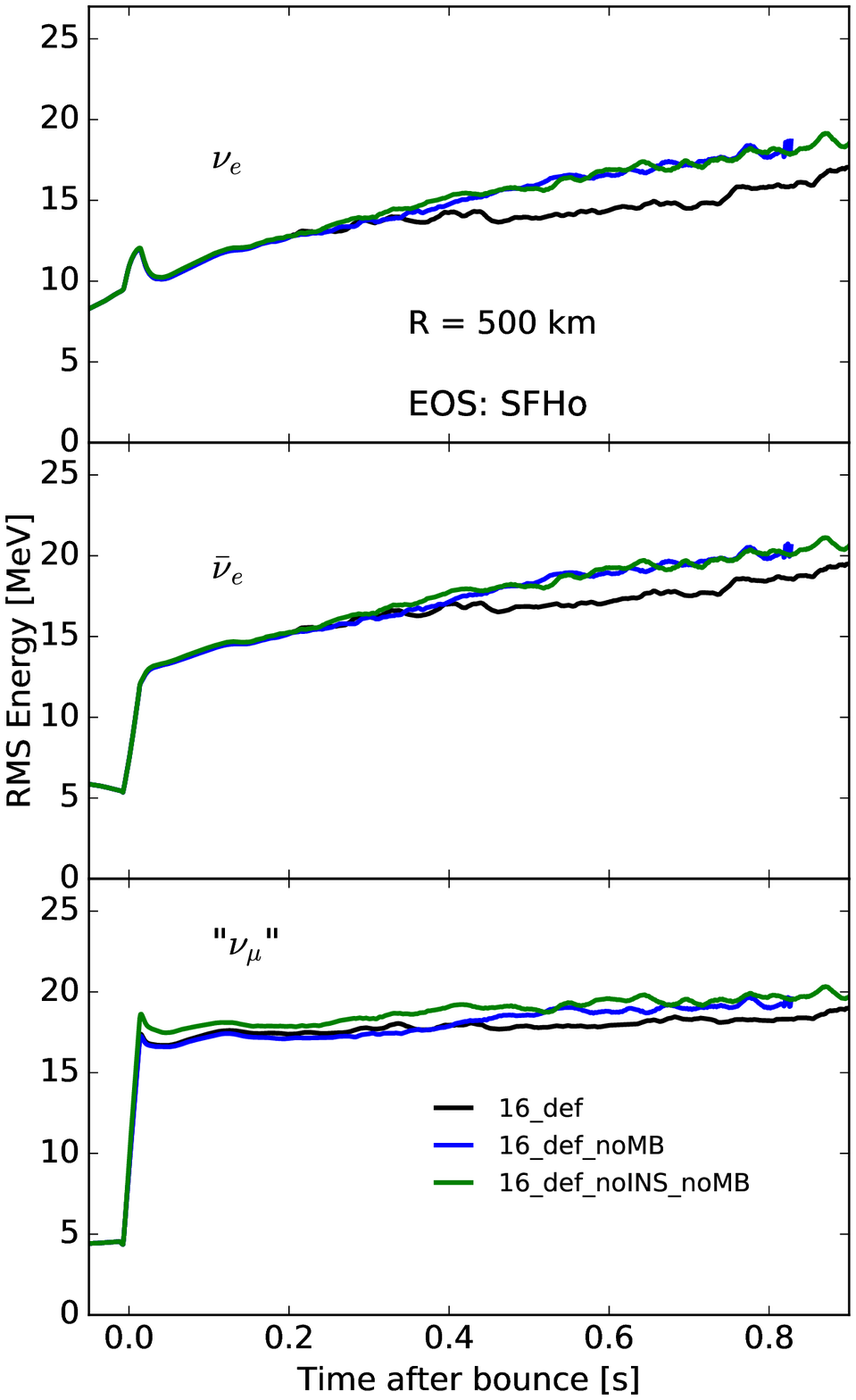}
\caption{Modification due to inelastic scattering off electrons and nucleons
of the luminosities (left) and RMS energies (right) of neutrinos at 500 km,
redshifted to the lab frame. Including inelastic scattering off nucleons
decreases the $\nu_\mu$ luminosities and RMS energies by $\sim$ 10\%, as
in M\"uller et al. (2012b), while slightly increasing the $\nu_e$ and $\bar{\nu}_e$
luminosities. RMS energies of the latter are mostly unaffected by inelastic
scattering. The default model (black, with many-body corrections and both
inelastic scatterings) shows a dip in luminosity and RMS energy after 300 ms
post-bounce, the time of its explosion.}
\label{fig:energy_micro}
\end{figure*}

Figure \ref{fig:spectra_micro} depicts the role of IES, INS, and MB on the 
emergent spectra at two different times after bounce. 
One of the central results that can be gleaned is the time of explosion.
This quantity can help one gauge the relative role of the inputs in question. 
The default model (def) here does explode. Prior to explosion, both inelastic scattering off
electrons and off nucleons and the many-body correction led to higher spectral fluxes.
After the default model explodes, the flux spectrum diminishes relative to
that of the non-exploding model 16$_{\_}$def$_{\_}$noMB. The slight increase in 
the heating in the gain region prior to explosion due to the inclusion of  
inelastic neutrino-electron and neutrino-nucleon scattering has certainly helped the 
core achieve the critical explosion condition. Figure \ref{fig:energy_micro} compares the
emergent neutrino luminosities (left panel) and root-mean-square (rms) neutrino energies (right)
for three models. The boosting in the $\nu_{\mu}$ luminosity and rms energy is clearly manifest,
as are the corresponding boosts in those quantities for the $\nu_e$ and $\bar{\nu}_{e}$ neutrinos.
While one may have speculated that enhanced neutrino losses, particularly due to $\nu_{\mu}$s
that are almost ineffectual in heating the shocked mantle, would have had a negative effect
on explodability, the converse is true.  Greater losses lead to a further compaction
of the core with an increase in the matter temperatures near the $\nu_e$ and $\bar{\nu}_{e}$ 
neutrinospheres. The result is similar in effect to that of GR, whereby such core heating enhances
the driving $\nu_e$ and $\bar{\nu}_{e}$ luminosities and the average neutrino energies, which in turn 
enhance the heating power in the gain region.  Since it is this power deposition that ultimately 
drives explosions, the net effect is quite supportive of explosion.  

However, the actual magnitude and form of the correction to the axial-vector coupling 
term in the expression for the neutrino-nucleon scattering rate may be different 
from that derived by Horowitz et al. (2017) and this still needs to be verified.  Moreover, the 
effects of similar many-body corrections to the absorption rates need to be incorporated,
as do those for the vector coupling strengths.  One prediction of the consequence
of the structure-factor correction we have employed is the enhanced $\nu_{\mu}$ luminosities
and average energies seen in Figure \ref{fig:energy_micro}.  It is noteworthy that, as it 
stands, the effect on the outcome of core-collapse of many-body rate suppressions 
(Burrows \& Sawyer 1998,1999) might be large.

\subsection{Equation of State}
\label{eos}

\begin{figure*}
\centering
\includegraphics[width=0.85\textwidth, angle = 0]{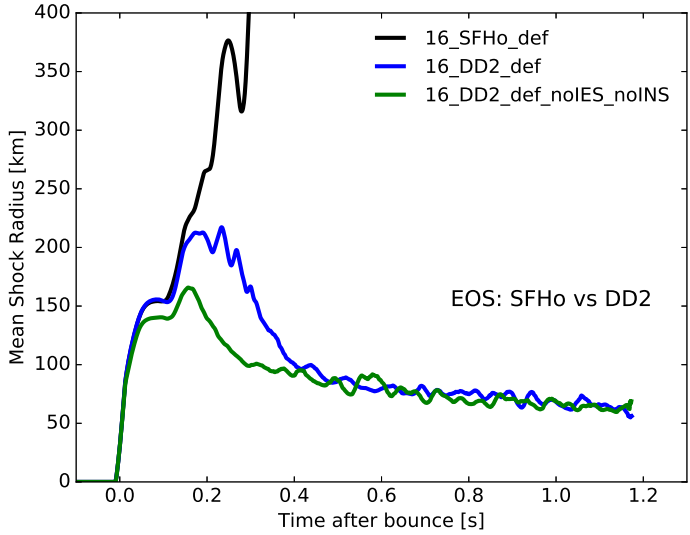}
\caption{Shock radii (km) versus time after bounce (s) for the SFHo and DD2 
EOS for the 16M$_{\odot}$ WH07 progenitor as a function of time after 
bounce. Only the former (our default model) explodes. We also plot for 
comparison a model with the DD2 EOS, but without any inelastic scattering 
off electrons or nucleons. Though neither DD2 model explodes, including 
inelastic scattering increases the stalled shock radius by $\sim$ 70 km.}
\label{fig:rshock_eos}
\end{figure*}

Through its control of the pressure for a given temperature, density, and Y$_e$, 
the equation of state of hot, lepton-rich nuclear matter will determine the structure
of the proto-neutron star and its evolution after bounce.  The stiffer the EOS,
the more extended the core.  On the one hand, a stiff EOS will resist
the quick increase in temperature near the $\nu_e$ and $\bar{\nu}_e$ neutrinospheres, 
and the consequent enhancement of the neutrino heating rate in the gain region,
seen both in the comparison of GR and Newtonian models and in the increase in the
$\nu_{\mu}$ losses due to the many-body effect. On the other, a stiffer EOS will 
provide a more stable inner platform that won't as easily (by its inward motion 
with time after bounce) send out weakening rarefactions to the outer bounce shock
that could inhibit explosion. Figure \ref{fig:rshock_eos} indicates (at least for this
16-M$_{\odot}$ model) that the first effect seems to win, since the DD2 EOS model 
does not explode, while the SFHo model (the softer of the two at high densities) 
does. On this figure we also show that, though the DD2 model didn't explode, 
the behavior of the shock with time was significantly less vigorous when the inelastic 
scattering effects were turned off, reiterating the conclusions of \S\ref{inelastic} and 
\S\ref{inel}. We also note that since the shock in the default DD2 model achieved a large 
mean shock radius ($\sim$200 km) before subsiding, this model was nevertheless very 
close to exploding.

\subsection{Nucleon-Nucleon Bremsstrahlung and Electron-Capture on Heavies}
\label{brems}

On Figure \ref{fig:rshock_full}, we also provide a model that drops the
many-body correction to neutrino-nucleon scattering, as derived by Horowitz et al. (2017),
but substitutes in the Fischer (2016) correction to the nucleon-nucleon bremsstrahlung rate
(``bf") and divides the electron capture rate on heavy nuclei derived for 
the SFHo EOS by Juodagalvis et al. (2010) by five (``0.2j").
The former effect lowers the bremsstrahling production rate of $\nu_{\mu}$ neutrinos, 
as well as the inverse bremsstrahlung absorption.  The latter effect slightly retards the 
infall rate, thereby decreasing the mass accretion rate post-bounce at a 
given time.  Decreasing this rate can faciliate explosion.  As the figure suggests,
these microphysical changes can compensate for ignoring the many-body effect
to result in a similarly-aided explosion.  Clearly, such modest alterations in the
microphysics of relevance, still within the envelope of our ignorance, could
play a positive role.  In Figure \ref{fig:rshock_15} below, we see a similar positive effect
of these potential changes in the default microphysics for a 13-M$_{\odot}$ model.
Together with the model in Figure \ref{fig:rshock_full}, this suggests that a more 
extensive exploration of such physics would be fruitful.

\section{Progenitor Perturbations}
\label{perturb}

Performing the last stages of stellar evolution before collapse hydrodynamically 
in 2D and 3D has been shown to alter, perhaps in significant ways, the
compositional, entropy, and density profiles of the core (Meakin et al. 2011; Couch et
al. 2015; Chatzopoulos et al. 2016; M\"uller et al. 2016; Abdikamalov et al. 2016;
M\"uller et al. 2017), and it has long been known that progenitor density profiles have an 
impact on the outcome of collapse. This is implicit in the critical curve analysis
of Burrows \& Goshy (1993), where $\dot{M}$ and the accretion ram pressure 
play central roles.  It is also a factor in discussions
of the compactness parameter (O'Connor \& Ott 2011) and its extensions (Ertl et al. 2015).
In this vein, one notes that at least three groups (Kitaura et al. 2006; Burrows et al. 2007b; 
Radice et al. 2017) have already demonstrated that the 8.8-M$_{\odot}$ ``electron-capture" supernova
progenitor of Nomoto \& Hashimoto (1988), with its extremely steep density ledge, can explode in 1D by the
neutrino-driven wind mechanism, though the explosion energy is low ($\sim$$1-2\times$ 10$^{50}$ ergs).

However, when spherical progenitor models do not readily lead to explosion, the initial perturbation spectrum 
in the progenitor's convective silicon and oxygen zones could certainly affect 
the timescales for the generation of turbulence behind the stalled shock and 
be a factor in the onset of explosion (Couch \& Ott 2013; M\"uller 
\& Janka 2015; Couch et al. 2015; Abdikamalov et al. 2016; M\"uller et al. 2017).  
Specifically, the magnitude, character, and spectra of seed perturbations will 
affect how quickly turbulence reaches the non-linear regime and, perhaps, 
whether turbulence grows to non-linearity at all during the finite time the 
accreta are in the unstable gain region. 

Therefore, introducing physically-motivated 
seed perturbations that originate from and reflect the three-dimensional character
of the convective core of an actual massive star at its terminal stage of evolution is a
topic of some interest. However, most calculations done to date do not start with
true 3D convective structures, but with 1D models from the literature, and impose either ad hoc perturbations
in density or velocity randomly at the grid level or allow grid asphericities (such
as obtained when using a Cartesian grid) or truncation errors to act as seeds.  Neither of these approaches
is physical, and the resulting initial perturbation spectra lead to early growth
rates in the linear regime that reflect not the multi-D progenitor
perturbation structure, but the numerical development of convenient artificial
power spectra. This will affect how quickly turbulence reaches the non-linear
regime and, perhaps, whether turbulence grows to non-linearity
at all during the finite time the accreta are in the unstable gain region.
In addition, this may have a bearing on the post-bounce delay to a turbulence-aided 
explosion, with the consequent effect on the time and energy of that explosion.

Therefore, the seed perturbations that arise during oxygen and silicon burning 
prior to collapse might be key inputs into core-collapse
supernova theory, and Couch \& Ott (2013), M\"uller \& Janka (2015), Couch 
et al. (2015), M\"uller (2016), and M\"uller et al. (2017) have begun to explore this. However, mixing-length
theory, though inadequate as a comprehensive theory, still provides a measure of
the magnitude of velocity perturbations at the onset of collapse (M\"uller et al. 2016), and they are only a
few hundred to $\sim$500 km s$^{-1}$, with Mach numbers bounded by $\sim$0.08 (Woosley \& Heger 2007).  
This is not large.  

In this section, we provide a quick glimpse at the possible relative role of significant
perturbations on the timing and character of explosion in light of the other physics. 
To this end, we employ the methodology of M\"uller \& Janka (2015).
These authors impose a vector velocity perturbation map on their progenitor that 
is more realistic than many past approaches and renders a perturbation field that 
is independent of grid resolution. This (surprisingly) was rarely attempted in 
the past and provides a specific context for future comparison. 
We set the maximum perturbation speed on the grid to 1000 km s$^{-1}$, which as indicated 
earlier may be near or beyond the expected upper end of the range, a spherical harmonic index $\ell$ of 2, 
and a radial ``quantum number" $n$ of 5.  Both $\ell$ and $n$ are parameters in 
the M\"uller \& Janka (2015) formulation.  With this parameter set, we simulate 
in 2D the self-consistent multi-group evolution of 13- and 15-M$_{\odot}$ progenitors 
and compare the result to a default model for which the initial perturbations are 
much smaller and arose numerically from grid noise.

\begin{figure*}
\centering
\includegraphics[width=0.49\textwidth, angle = 0]{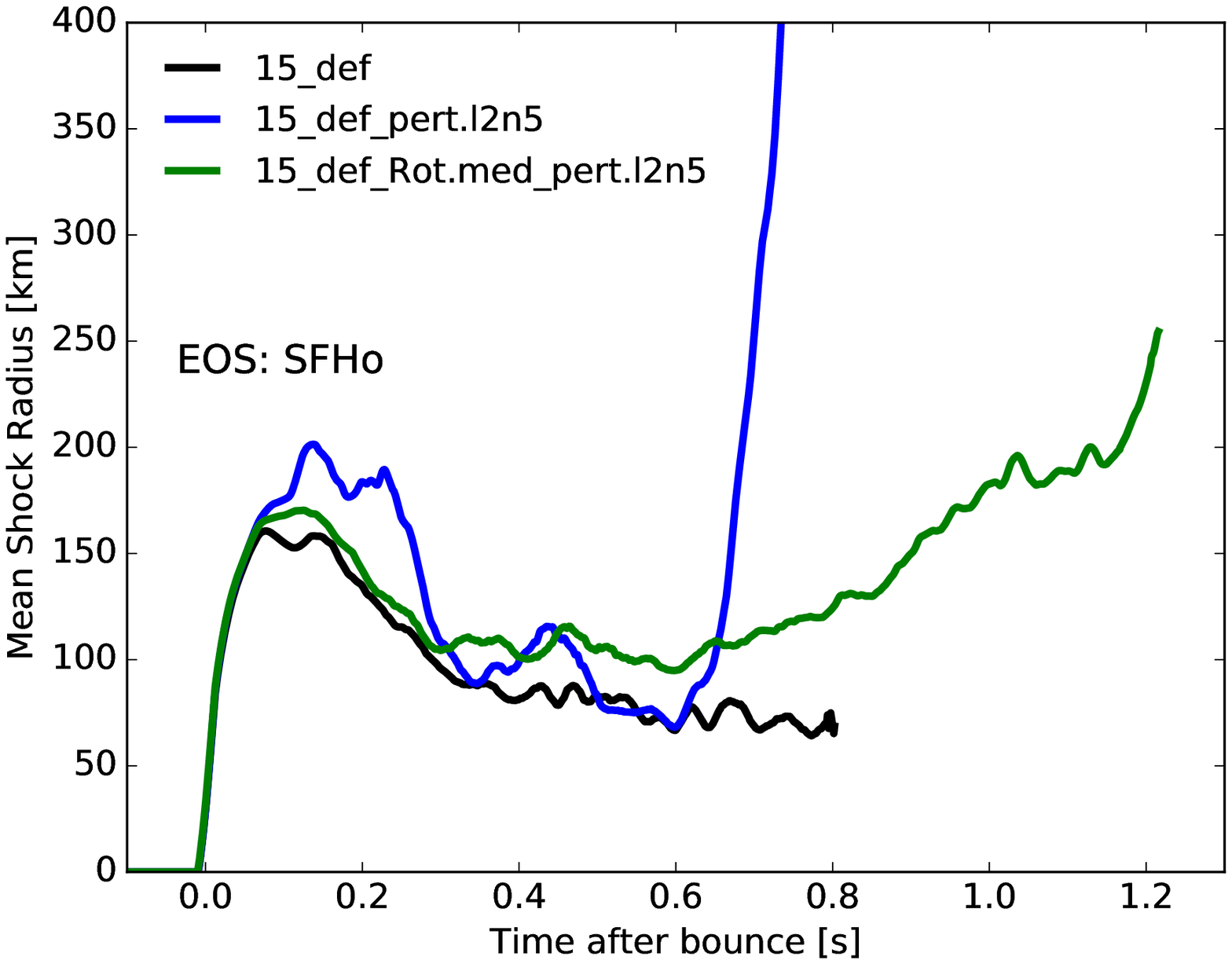}
\hfill
\includegraphics[width=0.49\textwidth, angle = 0]{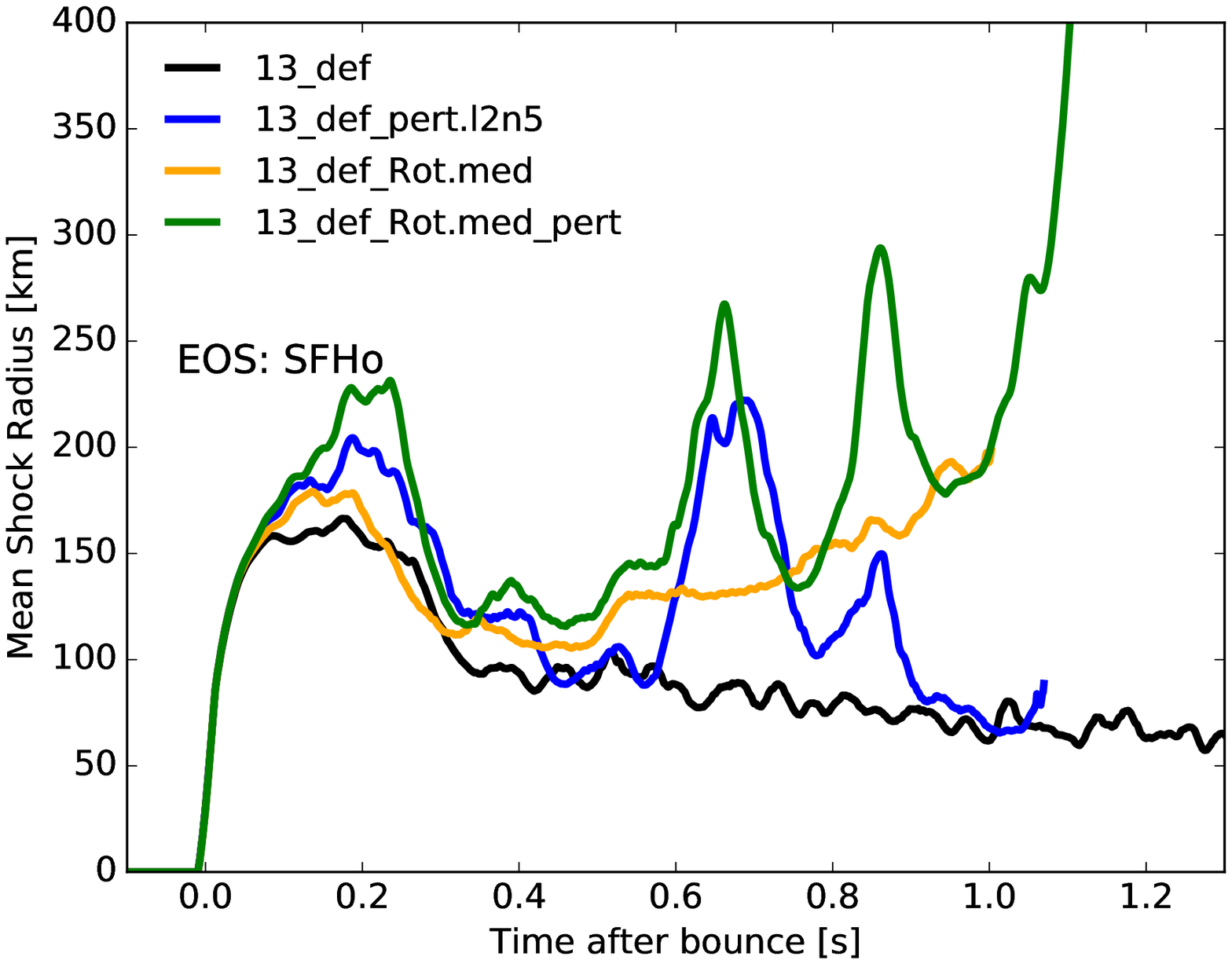}
\caption{Shock radius (in km) versus time after bounce (in s) for the 15 
(left) and 13 (right) M${\odot}$ Woosley \& Heger (2007) models, showcasing the effect of 
significant perturbations and moderate rotation.   We follow M\"uller \& Janka (2015)
prescription in implementing perturbations to radial velocities 
on infall over three regions with the maximal radial velocity (1000 km s$^{-1}$), 
$n$ (number of radial convective cells), and $l$ (number of angular convective cells) 
as parameters. The interior region spans 1000 to 2000 km, outside the nascent 
core, the middle region 2100 to 4000 km, and the outer region 4100 to 6000 km, 
truncated roughly where accretion ends after the first second for our simulations.  
All regions have $l=2, n=5$ and maximum radial velocity of 1000 km s$^{-1}$. 
See text for a discussion.}
\label{fig:rshock_15}
\end{figure*}

Figure \ref{fig:rshock_15} portrays the evolution of the mean shock radii, with and without 
progenitor perturbations.  The left panel shows that the imposed perturbations
converted failure into success for the 15-M$_{\odot}$.  We also provide on this panel 
a model including the effect of a modest rotation rate, along with the perturbation. 
Including the latter resulted in an explosion, though later.  
Our rotational study has revealed  that the effect
of rotation is not monotonic with initial internal rotational speed or spatial profile. 
On the right panel of Figure \ref{fig:rshock_15}, we show the corresponding behavior for
the 13-M$_{\odot}$ progenitor, which does not explode in our study for the default
microphysics.  It does, however, explode when similar perturbations are imposed
and when a similar rotational profile is assumed.  In this case, rotation promotes
explosion, highlighting its non-monotonic effect upon outcome. 
%

We note that the initial perturbations we imposed have an amplitude that 
is somewhat larger than expected (M\"uller et al. 2016) and that their character is 
still rather artificial. The magnitude of the initial perturbations may well be lower, but
their character and magnitude will certainly vary from progenitor to 
progenitor. Therefore, a much more thorough study with 3D progenitors 
and 3D collapse models is called for.

\section{Ray-by-ray+ Anomalies}
\label{ray}

As shown by Skinner et al. (2016), the ray-by-ray+ approach to
neutrino transport, whereby multi-D transport is replaced by multiple 1D transport 
calculations with corrections for matter advection, but not lateral transport, 
can introduce systematic errors in the heating rates along the poles
in axisymmetric 2D simulations. Such enhancements, when in proximity to criticality,
may be producing explosions artificially. At the very least, the time to explosion is 
artificially shortened, perhaps significantly. Since there is little or no accumulation of explosion 
energy prior to global instability (Burrows, Hayes, \& Fryxell 1995),\footnote{Only
those neutrinos emitted after the onset of the explosion contribute to                           
the asymptotic explosion energy.} an earlier 
explosion may make more of the emitted neutrinos available for explosive driving.

\begin{figure*}
\includegraphics[width=0.85\textwidth, angle = 0]{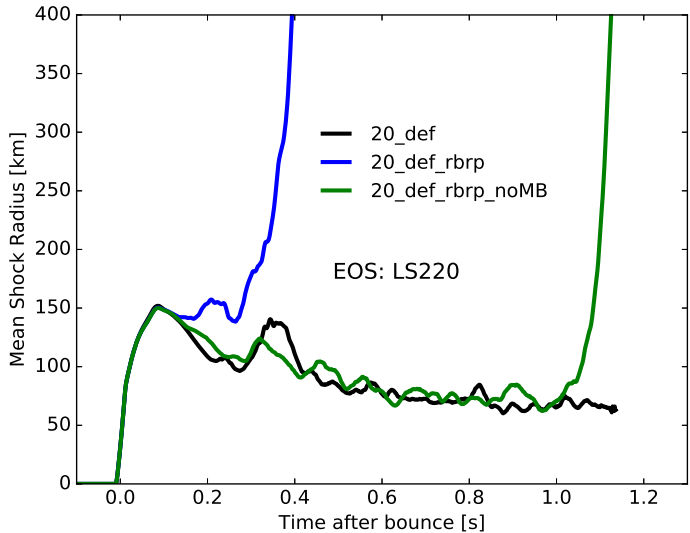}
\caption{Shock radius (km) versus time after bounce (s) for the 20 M${_\odot}$ 
WH07 progenitor using the LS220 EOS, with and without the ray-by-ray-plus (rbrp) 
approximation to neutrino transport. All models include inelastic scattering off 
electrons and nucleons. We see that including ray-by-ray+ (rbrp) leads to an explosion when 
otherwise there was none. The model with ray-by-ray+ and without many-body (green) explodes as well, though
700 ms after the model with ray-by-ray+ and many-body, suggesting that the ray-by-ray+, though artificial, is more significant to explosion than the physical
inclusion of the many-body correction. 
}
\label{fig:rshock_rbrp}
\end{figure*}

Figure \ref{fig:rshock_rbrp} compares three 20-M$_{\odot}$ models, using this time the 
LS220 EOS. Here, the default model does not explode, but the one employing the ray-by-ray+ 
simplification does.  
Note that, for both ray-by-ray+ models, the model without the many-body (MB) correction 
explodes much later (by 700 ms) than the model with MB. Our results suggest, for this
progenitor, that the artificial ray-by-ray+ approximation is more significant to explosion than the physical many-body effect.
One is left to speculate whether 2D simulations in the literature that 
employ ray-by-ray+ would indeed explode if they used more realistic transport.  
This is all the more relevant in 3D, given that extant published models 
that do explode in 2D have more difficulty exploding in 3D,
a context in which it is not clear that the ray-by-ray+ method introduces as great 
an artifact as in 2D.  It is true that the turbulent pressure spectra in 2D and 3D
are different, with the turbulent cascade in 2D resulting in enhanced stresses on larger scales, 
and that the turbulent-stress boost to explodability may be smaller
in 3D. This could also be a factor in the more anemic outcomes in published 3D 
models.  Nevertheless, it may be that the more problematic nature of published
3D models vis-\`a-vis 2D models is a consequence of some combination of the use
of ray-by-ray+ and the reduced turbulent stress in 3D.
  
We conclude this section by noting that the calculations of Skinner et al. (2016) 
did not include various physical effects (such as inelasticities and the many-body correction)
that we highlight here.  In Skinner et al., we obtained explosions using
ray-by-ray+ for some models that did not explode otherwise, and when they exploded
they did so late.

\begin{figure}
\includegraphics[width=\textwidth, angle = 0]{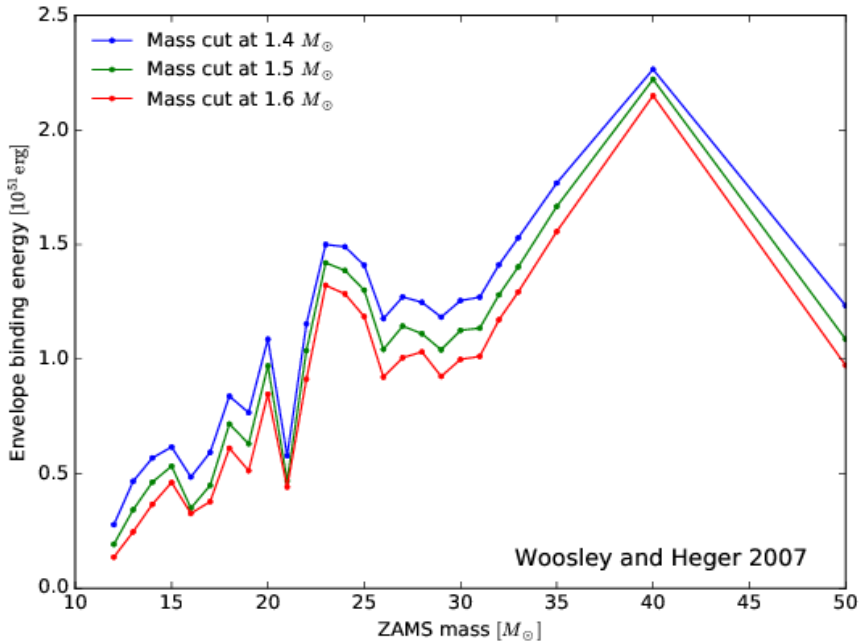}  
\caption{The outer envelope binding energies (in Bethes [$\equiv$ 10$^{51}$ erg])
for baryon mass cuts of 1.4, 1.5, and 1.6 M$_{\odot}$ versus ZAMS mass (in M$_{\odot}$) for
the Woosley \& Heger (2007) progenitor models. 
\label{ebind}}
\end{figure}

\section{Compactness}
\label{compact}

Finally, we conclude that although the compactness parameter (O'Connor \& Ott 2011,2013; Pejcha \& Thompson 2015), 
defined at bounce for a given mass interior ($M$) at a given radius ($R$) as 
$\xi \equiv M/M_{\odot}(1000{\rm km}/R)$, is one measure of the 
important density structure of the progenitor, it is not necessarily predictive 
of explosion, at least during the first second after bounce. We find that, 
depending upon the neutrino physics employed, the temporal order in which 
models with different compactness parameters explode after bounce varies.  
{If time of explosion is a fit measure of explodability, then this alone 
would challenge the usefulness of the compactness concept vis \`a vis explodability.}
Nevertheless, though the compactness parameter is large for the 21-M$_{\odot}$ model
and small for the 12-M$_{\odot}$, we found that the former can be more
explodable. One might have thought that if compactness were the sole predictor
the details of the neutrino-matter interaction would have mattered 
little in this regard. However, we see that the time to explosion does not 
necessarily correlate well with compactness.  Models with dense envelopes
have larger accretion and total neutrino luminosities after bounce,
and it is these luminosities that drive explosion. In addition, models
with dense envelopes and higher compactness have a greater optical depth 
to neutrinos in the gain region.  Since the neutrino power deposition goes
approximately as the product of this depth and the luminosities, 
high-compactness progenitors have an advantage.  Therefore, it is feasible
that more massive models might be more explosive, and this
trend has been seen by other modelers (Summa et al. 2016; Bruenn et al. 2016).
In fact, in the Summa et al. and Bruenn et al. papers for the Woosley \& Heger
(2007) models, the post-bounce explosion times increase in the sequence
20, 25, 15, and 12 M$_{\odot}$, which is clearly not monotonic with compactness.
This behavior is the reverse of what might be expected if low compactness 
signaled greater explodability.  Finally, the work of Nakamura et al. (2015)
suggests that though they see a weak correlation, it is not monotonic
with compactness. In fact, many of their plots versus compactness resemble 
scatter plots.

\begin{figure}
\includegraphics[width=\textwidth, angle = 0]{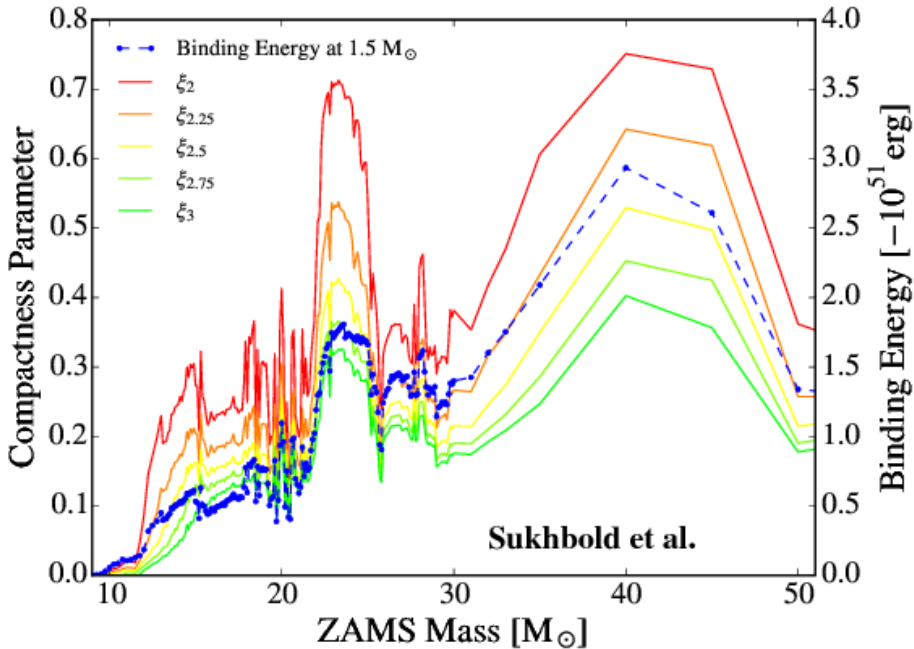}
\caption{This plot depicts the dependence of the compactness parameter (O'Connor \& Ott 2011,2013),
calculated at various interior masses, versus progenitor ZAMS mass, as well as the corresponding
envelope binding energy (blue dots; in Bethes [10$^{51}$ ergs]) for a baryon mass cut of 1.5 M$_{\odot}$
(see Figure \ref{ebind}).  The progenitor models of Sukhbold \& Woosley (2014) (also in
Sukhbold et al. 2016) are used. As this figure shows, whatever the position at which compactness is defined,
it correlates extremely well with envelope binding energy.  It is our contention that it
is the latter quantity that is more germane to the outcome of core collapse.
\label{compactness}}
\end{figure}

A critical issue is whether these more massive models with shallower density
profiles that explode early can maintain explosion during the traversal of
the shock through the outer stellar mantle.  The binding energy penalty
of the outer envelope generally increases with progenitor mass and might be too 
steep a price to pay during subsequent evolution for those more massive cores that
explode earlier and, perhaps, more energetically. As previously stated, 
for most of the relevant mass function, the envelope binding energy 
exterior to a given interior mass is an increasing function of progenitor 
mass.  It is this ``barrier" that may set the limit to the range of 
massive stars that can explode and leave behind neutron stars, 
although it cannot be excluded that the progenitor mass range 
that yields neutron stars, and not black holes, may be discontinuous 
(Sukhbold et al. 2016).  Figure \ref{ebind} portrays various exterior binding 
energies and Figure \ref{compactness} follows with a depiction of the close
correspondence between the compactness and the envelope binding energy 
exterior to a baryonic mass cut of 1.5 M$_{\odot}$. Therefore, we contend 
that whatever significance there may be to the compactness parameter is likely due 
to its correspondence with the binding energy of the outer envelope exterior 
to a given mass cut and that compactness need not correlate with the apparent 
explodability during the first second after core bounce.  Importantly, 
the ultimate outcome will depend upon the progress of the shock at 
post-bounce times that generally exceed those to which most core-collapse 
simulations currently go.

\section{Conclusions}
\label{conclusions}

In this paper, we have generated and explored detailed 2D (axisymmetric) {approximate} GR models of core-collapse 
supernovae using the new code F{\sc{ornax}}, treating all the relevant physics
to determine the dependence of the mechanism of explosion and its timing
on the physical and numerical inputs and assumptions. These include inelastic 
neutrino-electron and neutrino-nucleon scattering, many-body/structure-factor corrections
to the neutrino-nucleon scattering rates, nucleon-nucleon bremsstrahlung, 
electron capture on heavies, and physically-motivated initial perturbations.  We have 
also reexamined in brief the effect of using the ray-by-ray+ simplification to neutrino transport.
We found that much of the wide variation between the results obtained by different groups
around the world simulating stellar collapse (as well as whether the core explodes 
at all) might be explained by slight variations at the $\sim$10-30\% level in the 
microphysical inputs when the models are near the critical condition for explosion. 
In the process, we gauged the relative importance of otherwise sub-dominant neutrino
physics processes to the outcome of collapse. Proximity to criticality amplifies the dependence upon 
small changes in the neutrino sector that translate into slight, but crucial, changes 
in the emergent luminosities and average neutrino energies and the consequent 
post-shock heating rates; such sensitivity is not manifest in 1D simulations 
for which the core is (most often) far from explosive.  

Thus, ``Mazurek's Law" of severe feedback under variations in neutrino cross 
sections and rates is overturned due to the proximity to instability 
possible in the realistic multi-D turbulent context.  
While alterations in the neutrino coupling rates
have little effect in 1D, in multi-D the hydrodynamic response to even small changes and/or
corrections to neutrino interaction rates can be more substantial due
to greater proximity to the critical curve. 

The upshot is that small variations between the methods,
microphysics, and resolutions used by groups who ostensibly are incorporating the ``same" 
inputs translates naturally into post-bounce explosion time differences that can range
by many hundreds of milliseconds, and in some cases can turn a dud into an explosion (or vice versa).
We suggest that this thereby explains in large measure the apparent heterogeneity in 
the outcomes of detailed simulations performed internationally. A natural conclusion is that, 
viewed correctly, the different groups are collectively closer to a realistic understanding of 
the neutrino-driven mechanism of supernova explosion than might have seemed apparent 
and that a push to rationalize approaches and understand microphysical details 
in the neutrino-matter interaction sector and the nuclear equation of state 
could bring a resolution to the decades-long quest for a predictive model 
of core-collapse supernova explosions.   

We have found that many-body corrections to neutrino-matter interaction rates, even at 
sub-nuclear densities, can have similar effects as general relativity, progenitor 
perturbations, or inelastic neutrino-electron and neutrino-nucleon scattering.  
However, we caution that the actual magnitude and functional form of all 
the various many-body corrections to the neutrino-matter rates (both neutral- and 
charged-current), however important they seem from our current simulations, 
still need to be explored and verified. 

We performed a test using the ray-by-ray+ approximation to neutrino transport in a manner
similar to that employed by Skinner et al. (2016) to gauge its effect on the outcome
of collapse when a full physics suite is employed.  In Skinner et al. (2016), it was shown
that ray-by-ray+ artificially enhanced heating along the poles in synchrony with
the axial sloshing seen in 2D simulations and thereby made models more explosive.
The shift in the explosion times can be as large as the full range currently witnessed by the 
various supernova simulation groups for a given progenitor.  One can speculate 
that had groups that use the ray-by-ray+ simplification used real multi-D transport
their 2D models would have exploded later, perhaps much later or not at all.  Speculating 
further, one wonders whether the fact that 3D models explode later than
the corresponding 2D models (Lentz et al. 2015) or not at all (Melson et al. 2015, if
without their strangeness fix) may be connected with their use of ray-by-ray+, 
since non-rotating 3D simulations seldom manifest the axial sloshing seen in 2D.  
In short, without ray-by-ray+, it is not clear that 2D would explode much earlier than 3D.
However, the differences between the 2D and 3D convective cascades may be equally in play here and 
the associated simulations still need to be performed to assess this. 

The possible role of initial perturbations as seeds to the growth of 
convective instability in and around the gain region has been a subject
of recent focus. Clearly, allowing grid noise, truncation error, or 
other numerical noise to initiate linear growth to the non-linear phase,
or imposing artificial initial perturbations, is less than satisfactory.
This is particularly true given that seeds have a finite time to grow 
after accreting through the shock before leaving the unstable region
and given that the time to instability and explosion is germane to 
which phase of the neutrino light curve is driving explosion.  
Inaugurating the non-linear convective phase early and maintaining it
until explosion may be important, and the magnitude and timing of 
convective seeding is therefore of interest.  Though we have deferred 
until later a more comprehensive study of this subject, we tested the effect
of adding to the progenitor velocity perturbations whose magnitude was 
informed by mixing-length theory (M\"uller et al. 2016). In fact, 
we allowed the maximum perturbation speed to be 1000 km s$^{-1}$, 
with a Mach number as high as $\sim$0.12, which is a bit larger than found 
in 1D progenitor models (cf. Woosley \& Heger 2007).  Indeed, we found that 
large imposed perturbations of this sort could enable explosion, though 
the magnitude of the imposed perturbation seems large. 
Clearly, it will be important to determine their character, and the 
many 3D-progenitor studies now in process promise to do just that.  
 
Many-body corrections to scattering rates, inelastic scattering, decreases
in nucleon-nucleon bremstrahlung rates and in electron capture rates on heavy nuclei,
and initial seed perturbations all boost ``explodability" and shorten the time 
to explosion.  In fact, these effects add synergistically and non-linearly 
to aide explosion, despite the fact that they individually amount to effects 
at the $\sim$10$-$30\% level. This is due to the proximity of multi-D models 
to criticality, and is not seen in 1D simulations.

Furthermore, the turbulence behind the shock that has been shown to aid explosion
in the realistic multi-D context naturally introduces indeterminacy in detail.
Even the same progenitor star, but with different random seed perturbations and rotational structures
at collapse, should yield a range of explosion energies, nucleosynthesis, $^{56}$Ni yields,
pulsar kicks, and explosion morphologies.  Therefore, it is expected that Nature
provides distribution functions, and not one-to-one maps, in all signatures of explosion.
In the long run, theory will need to come to grips with this, but in the short run one 
should not expect that in the chaotic context of turbulent convection and pre-collapse
structures the best models will correspond in detail. This is the physical and natural
consequence of turbulence and chaos, and will be paralleled in comparison verification studies.

The next stage is to explore the same issues in three dimensions, 
and one expects there to be important differences (Takiwaki et al. 2014; 
Lentz et al. 2015; M\"uller 2015; M\"uller et al. 2017). It is only 
after performing such simulations, and their subsequent verification, that a robust 
resolution to the core-collapse supernova problem can be claimed.  
Nevertheless, there has been significant progress of late in unraveling
this central mystery in astrophysics, in demonstrating the viability of
the neutrino-driven mechanism of explosion, and in illuminating 
its component physics.

\begin{acknowledgements}
The authors acknowledge the help of Evan O'Connor with the Lattimer-Swesty
equation of state and of Todd Thompson, who was instrumental in
developing the inelastic scattering tables and schema.
In addition, they thank Chuck Horowitz for early conversations concerning
the neutrino response in nuclear matter at low densities, his insights into
neutrino-matter interaction physics, and for an advanced look at his
recent paper on these topics. Finally, they would like to thank Yukiya
Saito and Junichiro Iwasawa for help scrutinizing the various progenitor
model sets in the literature and Sean Couch for stimulating conversations
on a variety of core-collapse topics.
Support was provided by the Max-Planck/Princeton Center 
(MPPC) for Plasma Physics (NSF PHY-1144374) and NSF grant AST-1714267. 
The authors employed computational resources provided by the TIGRESS
high performance computer center at Princeton University, which is jointly supported by the Princeton
Institute for Computational Science and Engineering (PICSciE) and the Princeton University Office of
Information Technology, and by the National Energy Research Scientific Computing Center
(NERSC), which is supported by the Office of Science of the US Department of
Energy (DOE) under contract DE-AC03-76SF00098. The authors express their gratitude to
Ted Barnes of the DOE Office of Nuclear Physics for facilitating their use of NERSC.
This work was originally part of the ``Three Dimensional Modeling of Core-Collapse
Supernovae" PRAC allocation support by the National Science Foundation (award number ACI-1440032)
and was part of the Blue Waters sustained-petascale computing project,
which is supported by the National Science Foundation
(awards OCI-0725070 and ACI-1238993) and the state of Illinois.
Blue Waters is a joint effort of the University of Illinois at
Urbana-Champaign and its National Center for Supercomputing Applications.
This paper has been assigned a LANL preprint \# LA-UR-16-28849.
\end{acknowledgements}







\end{document}